\documentclass[twocolumn,showpacs,amsmath,amssymb,prb,superscriptaddress]{revtex4-1}
\usepackage{graphicx}
\usepackage{color}
\begin{document}

\title{Wigner crystal in a two-dimensional electron system
 in the vicinity of filling factor 1/5: Acoustic studies}

\author{I.~L.~Drichko}
\author{I.~Yu.~Smirnov}
\affiliation{A.~F.~Ioffe Physico-Technical Institute of Russian
Academy of Sciences, 194021 St. Petersburg, Russia}
\author{A.~V.~Suslov}
\affiliation{National High Magnetic Field Laboratory, Tallahassee,
FL 32310, USA}
\author{L.~N.~Pfeiffer}
\author{K.~W.~West}
\affiliation{Department of Electrical Engineering, Princeton
University, Princeton, NJ 08544, USA}
\author{Y.~M.~Galperin}
\affiliation{Department of Physics, University of Oslo, 0316 Oslo,
Norway} \affiliation{A.~F.~Ioffe Physico-Technical Institute of
Russian Academy of Sciences, 194021 St. Petersburg, Russia}

\begin{abstract}
By simultaneous measurements of the attenuation and velocity of
surface acoustic waves propagating in proximity to a high-quality
GaAs quantum well we study the complex AC conductance of the
two-dimensional electron system. Focusing on the vicinity of the
filling factor $\nu=1/5$ we confirm that the insulating states
formed closely to this value of $\nu$ are pinned Wigner crystals.\\[0.1in]
\end{abstract}


\maketitle

\section{Introduction}

The nature of the ground state of a two-dimensional electron system
(2DES)  in a large perpendicular magnetic field $B$ has attracted a
lot of attention. At small filling factors, $\nu=2\pi \hbar n/eB $
where $n$ is the 2DES density, $\hbar$ is the reduced Planck
constant, and $e$ is the electronic charge, the ground state in the
absence of disorder is expected to be the Wigner crystal
(WC)~\cite{Lozovik1975,Yoshioka1979,Fisher1982,Yoshioka1983,Lam1985}.
Another known ground states are the fractional quantum Hall effect
(FQHE) states~\cite{Tsui1982,Laughlin1983}. Both states are induced
by the electron-electron interaction. It turns out that the Laughlin
FQHE liquid states at $\nu = p/q$ (where $p$ and $q$ are integers)
are particularly robust and have ground state energies which are
lower than the WC state energy, at least for $\nu
>1/5$~\cite{Sajoto1993}. Theoretical calculations predict that, in
an ideal 2DES system, the WC should be the ground state for $\nu$
smaller than about 1/6. However, the WC state may win as the filling
deviates slightly from 1/5. It is possible therefore to have a WC,
which is reentrant around a FQHE liquid state, see Fig.~9
in~[\onlinecite{Shayegan2005}]. This would rationalize the general
current belief that the insulating phase (IP) observed around the
$\nu=1/5$ FQHE in very high quality GaAs/AlGaAs 2DESs is the
signature of a WC state pinned by a disorder potential. The
magnetic-field-induced WC problem in 2DESs has been studied
extensively since the late 1980s~\cite{Shayegan1997,Pan2002}.

In 2D systems along with DC measurements of the components of the
magneto-resistance tensor many research groups study AC conductivity
$\sigma (\omega)$, which is a probeless way for investigation of the
AC conductance. The rf electric field can be excited using the
coplanar-wave-guide (CPW) technique~\cite{Wen1969}. This method was
successfully employed for studies of the FQHE
in~[\onlinecite{Engel1993}] and other works.

Another probeless method is using traveling electric wave created by
a surface acoustic wave (SAW). In connection with the integer QHE
structures it was implemented
in~[\onlinecite{Wixforth1986,Drichko2000}] and subsequent works; the
FQHE was studied using this method in~[\onlinecite{Paalanen1992}]
and [\onlinecite{Paalanen1992a}]. AC  methods provide the
information additional to the DC results. In particular, specific
resonances in the AC response allow one to identify the nature of
the insulating states observed at specific values of the filling
factor.

The microwave spectroscopy (MWS) based on the CPW technique and the
acoustic spectroscopy (AS) based on studies of the attenuation
$\Gamma$ and velocity \textit{v} of a SAW provide complementary
information. MWS allows studying the high-frequency response while
the frequency of the excited SAW is limited. However, the AS allows
calculating \textit{both} (real and imaginary) components of the
complex AC conductance from simultaneous measurement of the SAW
attenuation $\Gamma$ and variation of its velocity
$\Delta$\textit{v}/\textit{v}$_0$ versus perpendicular magnetic
field $B$. This is an obvious advantage of the AS, which we employ
in the present study.

Usually the frequency domains used in MWS and AS do not overlap.
However, recently is was shown that the frequency windows might
overlap providing a possibility for quantitative studies of AC
conductance in a broad frequency domain~\cite{Drichko2014}.

In the present work, we address the insulating phases observed in
the high-quality GaAs quantum wells in the vicinity of $\nu = 1/5$.
These phases are ascribed to formation of a disorder-pinned Wigner
crystal (WC)~\cite{Paalanen1992a}. Our aim is to extend studies of
this region by simultaneous analysis of the real and imaginary parts
of the complex AC conductance. Comparing the results with the
theory~\cite{Fogler2000} we show that the behaviors  of $\textrm{Re}
\sigma (\omega)$  and $\textrm{Im} \sigma (\omega)$ are indeed
characteristic for a pinned WC.

\section{Experiment}
\subsection{Methodology}

\begin{figure}[h!]
\centering
\includegraphics[height=5.5cm]{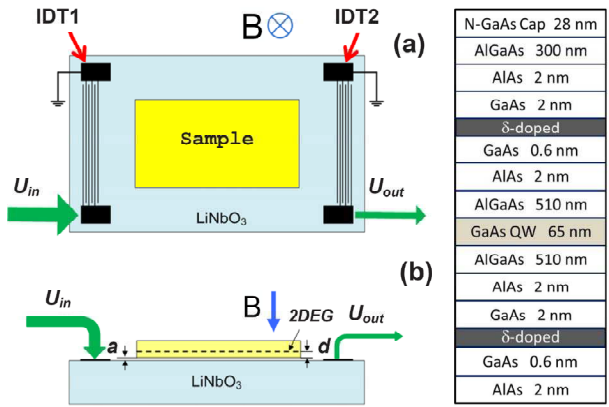}
\caption{Left panel: Sketch of the experimental setup; (a) top and (b) side views. Right panel: The structure of the sample.
 \label{fig1}}
\end{figure}

We use the so-called hybrid acoustic method discussed in detail in
Ref.~\onlinecite{Drichko2000}, see Fig.~\ref{fig1}~(left). The
sample is pressed by springs to a surface of a LiNbO$_3$
piezoelectric crystal where two inter-digital transducers (IDTs) are
formed. One of the IDTs is excited by AC pulses. As a result, a
surface acoustic wave (SAW) is generated, which propagates along the
surface of the piezocrystal. The piezoelectric field penetrates into
the sample interacting with the charge carriers. This interaction
causes SAW attenuation and deviation of its velocity. Measurement of
both SAW attenuation and velocity versus perpendicular magnetic
field allows one finding complex AC conductance, $\sigma (\omega)
\equiv \sigma_1(\omega) - i \sigma_2(\omega)$, as a function of
magnetic field.
\subsection{Samples}

We study multi-layered $n$-GaAlAs/GaAs/GaAlAs structures with a wide
(65 nm) GaAs quantum well (QW), see Fig.~\ref{fig1}~(right). The QW
is $\delta$-doped from both sides and is located at the depth $d$ of
845 nm from the surface. The electron density is $n=5\cdot
10^{10}$~cm$^{-2}$ and the mobility is $8\cdot
10^6$~cm$^2$/V$\cdot$s. One can expect that at the given electron
density only lowest band of transverse quantization is
occupied~\cite{Manoharan1996}.

\subsection{Experimental results}

Both attenuation of SAW and variation of its velocity were measured
in several samples versus perpendicular magnetic field of up to 18 T
in the frequency domain $f=(28.5-306)$~MHz  at temperatures
$T=(40-380)$~mK. The samples were cut form the same chip. The
electron density in the samples is diffed by $\approx 2$~\%.

\begin{figure}[ht]
\centering
\includegraphics[width=.75\columnwidth]{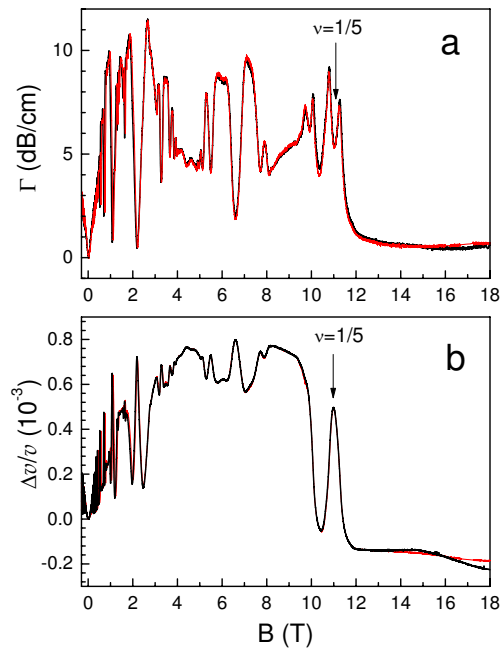}
\caption{ Magnetic field dependences of the (a)
attenuation $\Gamma$ and (b) relative deviation of the SAW velocity
$\Delta$\textit{v}/\textit{v}~(b). $f=142$~MHz, $T=45$~mK. Black curves correspond to increasing magnetic field while red ones correspond to decreasing field. \label{fig3}}
\end{figure}

Shown in Fig,~\ref{fig3} are magnetic field dependences of the
attenuation $\Gamma$ and relative deviation of the SAW velocity
$\Delta$\textit{v}/\textit{v} at $f=142$~MHz, and $T=45$~mK. The
curves for different  frequencies and temperatures are similar.
Magnetic field has first increased form 0 up to 18~T (black curves)
and then decreased (red curves) down to -0.3~T. The curves overlap
confirming that there is no hysteresis. We have checked that the
measurements correspond to the linear response regime.

The complex AC conductance $\sigma (\omega)$ was determined using
Eqs.~(1)-(7) from~\cite{Drichko2000} where we substituted
$\varepsilon_1=50$, $\varepsilon_0=1$ and $\varepsilon_s=12$ for the
dielectric constants of LiNbO$_3$, of the vacuum and of the sample,
respectively. The finite vacuum clearance $a = 5\cdot 10^{-5}$~cm
between the sample surface and the LiNbO$_3$ surface was determined
from saturation of the SAW velocity in a strong magnetic field at
$T=380$~mK; $d=845$~nm is  the finite distance between the sample
surface and the 2DES layer. The SAW velocity in zero magnetic field
is $v_0= 3\cdot 10^5$~cm/s.

\begin{figure}[b]
\centering
\includegraphics[width=.9\columnwidth]{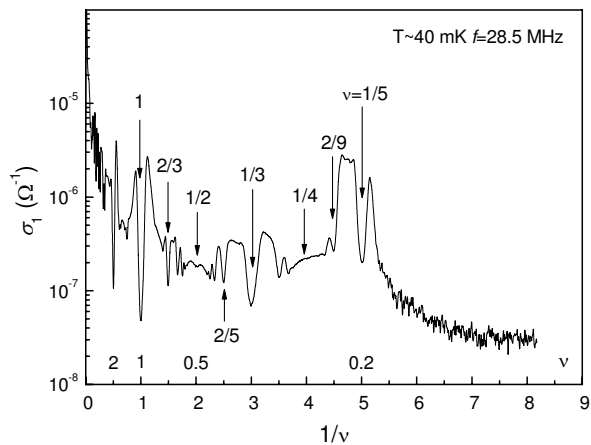}
\caption{Real part of the AC conductance,
$\sigma_1\equiv \textrm{Re} \sigma$, as a function of the inverse
filling factor, $\nu^{-1}$. \label{fig4}}
\end{figure}

Dependence of  $\sigma_1 \equiv \textrm{Re} \sigma$ on the inverse
filling factor $\sigma_1 \equiv \textrm{Re} \sigma$ is shown in
Fig.~\ref{fig4}. This picture evidences a rich oscillation pattern
including features of both integer and fractional quantum Hall
effect.

\begin{figure}[t]
\centering
\includegraphics[width=.75\columnwidth]{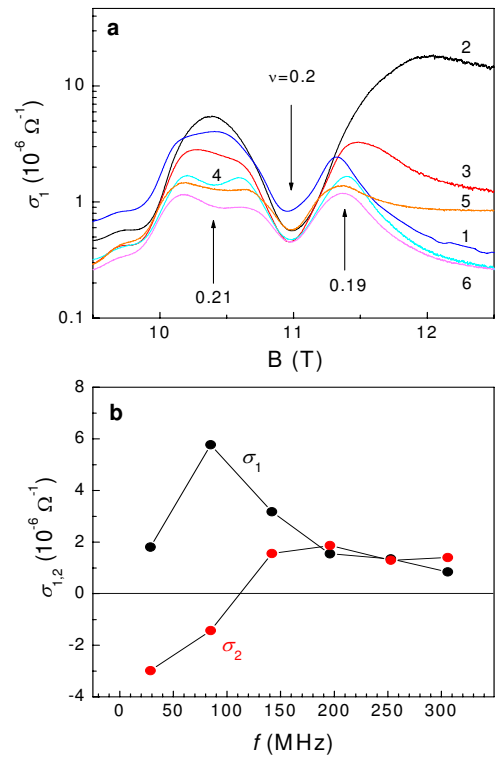}
\caption{(a)  Magnetic field dependences of $\sigma_{1}$ for different frequencies, MHz: 1-28.5, 2-85, 3-142, 4-191, 5-256, 6-306.
(b)  Frequency dependences of $\sigma_{1,2}$ for
$\nu = 0.19$, $T= 40$~mK. The lines are guides to the eye.
\label{fig5}}
\end{figure}

Our primary aim is to study $\sigma (\omega)$ in a vicinity of
$\nu=1/5$. The magnetic field dependence of $\sigma_1$ for different
frequencies is shown in Fig.~\ref{fig5}~(a). Frequency dependences
of $\sigma_1$ and $\sigma_2$ for  $\nu = 0.19$ and, $T= 40$~mK are
shown in Fig.~\ref{fig5}~(b).

As shown in Fig.~\ref{fig5}~(b), $\sigma_1(\omega)$ has a maximum at
$f\equiv \omega/2\pi \approx 85$~MHz, while the imaginary part,
$-\sigma_2 (\omega)$, changes its sign. Similar behaviors were
observed for $\nu=0.21$. A special behavior of $\sigma_1$ is
observed at $\nu = 1/5$. At this value of $\nu$  the frequency
dependence of $\sigma_1$ is smooth. This is compatible with the
conclusion that at $\nu=1/5$ the FQHE state wins and no Wigner
crystal is formed. The observed mode is close to the so-called
B-mode found in~[\onlinecite{Chen2004}] for a similar sample.

\subsection{Dependences on intensity and temperature}

\begin{figure}[h!]
\centering
\includegraphics[width=.75\columnwidth]{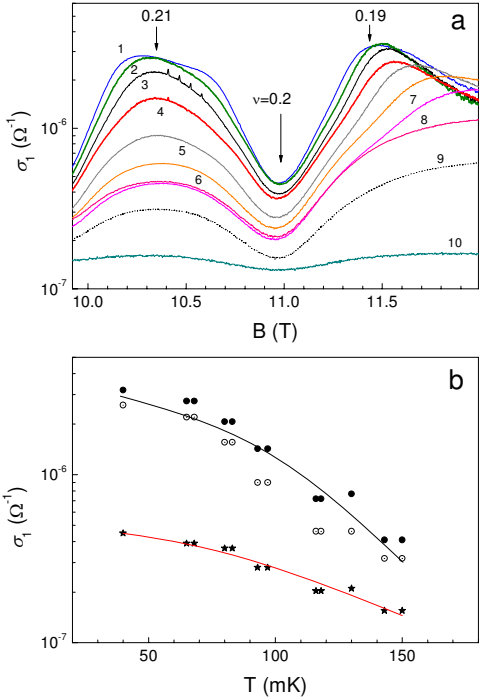}
\caption{(a)  Magnetic field dependences of $\sigma_{1}$ for different
temperatures, mK: 1-40, 2-45, 3-67, 4-82, 5-95, 6-105, 7-117, 8-130, 9-147,10-345.
(b)  Temperature dependences of $\sigma_{1}$ for $\nu = 0.19$
({\large{\textbullet}}), 021 ({\small{$\odot$}}) and 0.20 ($\bigstar$);
the lines are guides to the eye. $f=142$~MHz.
\label{fig6}}
\end{figure}

Magnetic field dependences of $\sigma_1$  at different temperatures
and its temperature dependences at different filling factors are
shown in Fig.~\ref{fig6}.
As temperature increases the maxima of $\sigma_1(B)$ (corresponding
to $\nu = 0.19$ and 0.21) decrease much more rapidly than the
conductance at $\nu = 1/5$. Following~[\onlinecite{Paalanen1992a}]
we interpret this behavior as melting of the Wigner crystal.

Note that temperature dependences of $\sigma_1 (\omega)$ are
qualitatively similar for $\nu=0.19$, 0.20, and 0.21, contrary to
the temperature dependences of DC conductance. Indeed, according
to~[\onlinecite{Jiang1990}] the temperature dependences of
$\rho_{xx}$ are \textit{qualitatively}  different for $\nu =0.20$
and $\nu=0.21$: compare Fig.~2 and Fig.~3 of that paper. At $\nu
=0.20$, $\rho_{xx}\propto  \exp\left [-1.1/T\, \text{(K)}\right]$
while at $\nu =0.21$, $\rho_{xx}\propto  \exp\left [0.63/T\,
\text{(K)}\right]$.

\begin{figure}[h]
\centering
\includegraphics[width=0.75\columnwidth]{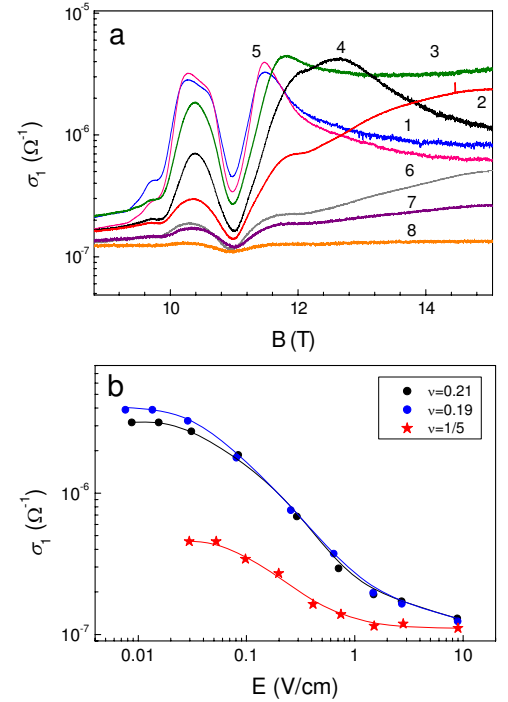}
\caption{(a) Dependence $\sigma_1(B)$ at different input
intensities, $\mu$W: 1 - $1.58\cdot 10^{-4}$,
2 - $1.58\cdot 10^{-3}$, 3 - $4.8\cdot 10^{-3}$,
4 - $1.6 \cdot 10^{-2}$, 5 -$ 4.9 \cdot 10^{-2}$,
6- $1.7\cdot10^{-1}$, 7- $6.2\cdot 10^{-1}$,
8 - $6.2\cdot 10^{0}$. $f=142$~MHz, $T=40$~mK.
(b) - Dependences of $\sigma_1$ versus SAW
 electric field amplitude for $\nu = 0.21$,  0.19 and 0.20; the lines are guides to the eye.
 The SAW amplitude $E$ was found
 by Eq.~(3) of~[\onlinecite{Drichko2013}].
\label{fi}}
\end{figure}

The dependences of $\sigma_1$ on magnetic field for different SAW
intensities $E$ are shown in Fig.~\ref{fi}~(a). Shown in
Fig.~\ref{fi}~(b) are the dependences of $\sigma_1$ versus the SAW
electric field  amplitude for different filling factros: $\nu =
0.19$, 0.20, and 0.21. These dependences were measured at frequency
$f=142$~MHz and temperature $T=40$~mK.
Since the dependences of $\sigma_1$ on temperature and SAW intensity
are qualitatively similar, one concludes that the main mechanism
behind nonlinear response to a SAW is heating of 2DES due to the
energy SAW dissipated in course of SAW attenuation.

\begin{figure}[h!]
\centering
\includegraphics[width=0.7\columnwidth]{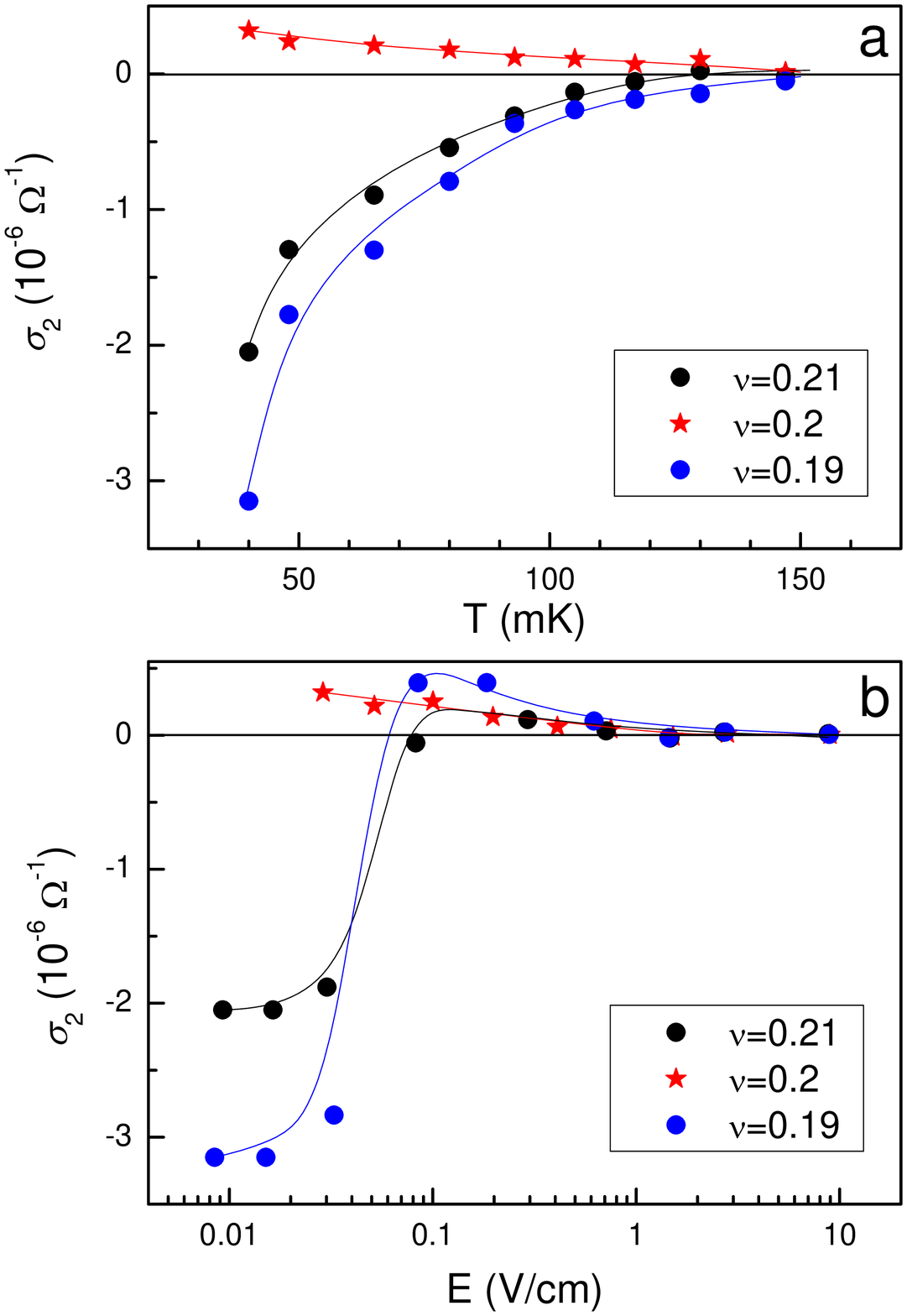}
\caption{ (a) Temperature dependence of $\sigma_2$ for $\nu = 0.19$, 0.2, and 0.21. $f=142$~MHz.
(b) Dependences $\sigma_2 (E)$ for $T=40$~mK. The lines are guides to the eye.
\label{ft}}
\end{figure}

Dependences of $\sigma_2$ on $T$ and $E$ at $f=142$~MHz are shown in
Fig.~\ref{ft}. Again, we observe that the temperature dependences of
$\sigma_2$ are similar for $\nu=0.19$ and $\nu=0.21$, but is
qualitatively different from the temperature dependence for
$\nu=0.20$.  The dependences of $\sigma_2$ on the amplitude of the
SAW electric field are similar to the temperature dependences, thus,
supporting the idea that the main mechanism behind nonlinear
behaviors is the Joule heating.

\section{Discussion and conclusions}

The behavior of $\sigma (\omega)$ shown in Fig.~\ref{fig6}~(b) is
typical for a pinned mode of a Wigner
crystal~\cite{Fertig1999,Yi2000,Fogler2000,Chitra2001,Fogler2004},
see~Refs.\onlinecite{Shayegan1997,Shayegan2005} for a review. It
manifests itself in observed  resonances in $\sigma_1
(\omega)$~\cite{Ye2002}, which has been taken as a signature  of a
solid and interpreted as due to the pinning mode (the disorder
gapped lower branch of the
magnetophonon)~\cite{Fukuyama1978,Normand1992,Fertig1999,Fogler2000,Chitra2001}
of WC crystalline domains oscillating collectively within the
disorder potential. The WC states compete with the fractional
quantum Hall effect (FQHE) states; based on several experiments and
calculations  it is concluded that at $\nu=1/5$ the FQHE dominates
while at $\nu$ slightly less or slightly higher the WC state wins,
see, e.g., Fig.~9 from~\cite{Shayegan2005}.

The dynamic response of a weakly pinned Wigner crystal at  not too
small frequencies is dominated by the collective
excitations~\cite{Fertig1999,Fogler2000,Fogler2004} where an
inhomogeneously broadened absorption line (the so-called pinning
mode) appears~\cite{Fukuyama1977,Fukuyama1978}. It corresponds to
collective vibrations of correlated segments of the Wigner crystal
around their equilibrium positions formed by the random pinning
potential. The mode is centered at some disorder- and
magnetic-field-dependent frequency, $\omega_p$ (so-called pinning
frequency); its width being determined by a complicated interplay
between different collective excitations in the Wigner crystal.
There are modes of two types: transverse (magnetophonons) and
longitudinal (magnetoplasmons). The latters include fluctuations in
electron density. An important point is that the pinning modifies
both modes, and the final result depends on the strength and the
correlation length, $\xi$, of the random potential. Depending in the
strength and  correlation length of the random potential, the
frequency, $\omega_p$ may either increase, or decrease with the
magnetic field.
The ratio $\omega_p/\omega_c$, where $\omega_c$ is the cyclotron frequency, can be arbitrary. Depending on the interplay between the ratio $\omega_p/\omega_c$ and the ratio $\eta \equiv
\sqrt{\lambda/\beta}$ between the shear ($\beta$) and bulk ($\lambda$) elastic moduli one can specify two regimes where the behaviors of $\sigma^{\text{AC}}$ are different:
\begin{equation} \label{regimes}
(a) \ 1 \ll \omega_c/\omega_{p0} \ll
\eta, \quad (b) \ 1 \ll \eta
  \ll \omega_c/\omega_{p0}     \, .
\end{equation}
Here $\omega_{p0}$ is the pinning frequency at $B=0$. As a result, the
variety of different behaviors is very rich.
 Assuming $\xi \gg
l_B=(\hbar c/eB)^{1/2}$ one can cast the expression for
$\sigma_{xx}(\omega)$ from Ref.~\cite{Fogler2000} into the form
\begin{equation}
  \sigma (\omega)=-i\frac{e^2n\omega}{m^*\omega_{p0}^2}\frac{1-iu(\omega)}
{[1-iu(\omega)]^2   -(\omega
  \omega_c/\omega_{p0}^2)^2}\, , \label{sigma1}
\end{equation}
where the function $u(\omega)$ is different for regimes (a) and (b).

Let us consider the regime (b) since only this regime seems to be
compatible with our experimental results. Then
\begin{equation}
  \label{eq:fb}
  u(\omega) \sim \left\{ \begin{array}{lll}
(\omega/\Omega)^{2s}, & \omega \ll \Omega\, , &\quad  (b1)\\
\text{const}, & \Omega \ll \omega \ll \omega_c \, .& \quad (b2) \end{array}
\right.
\end{equation}
Here $\Omega \sim \omega_{p0}^2\eta/\omega_c$, while $s$ is some
critical exponent. According to Ref.~\onlinecite{Fogler2000},
$s=3/2$.

\begin{figure}[ht]
\centering
\includegraphics[width=0.49\columnwidth]{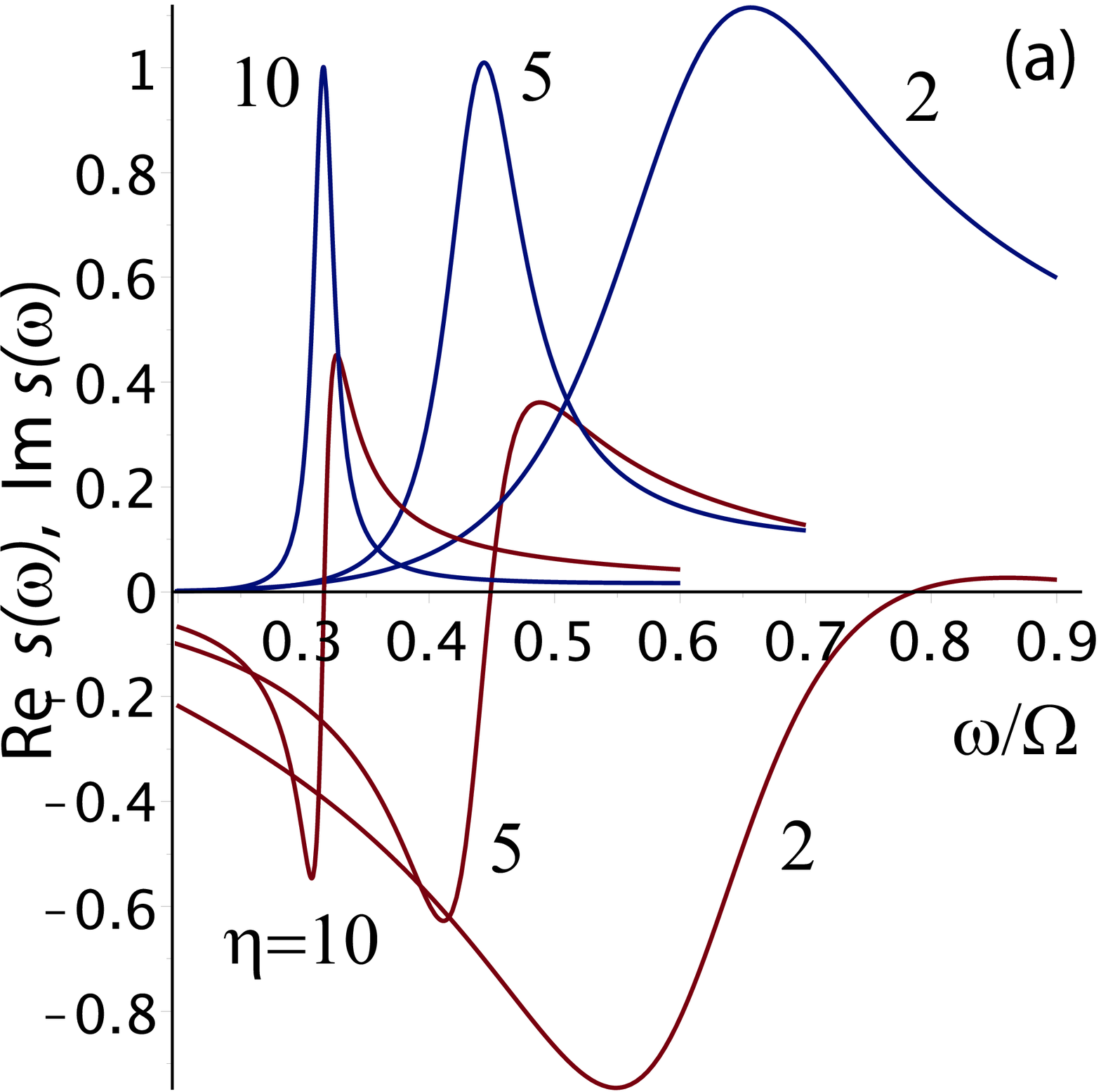} \hfill
\includegraphics[width=0.49\columnwidth]{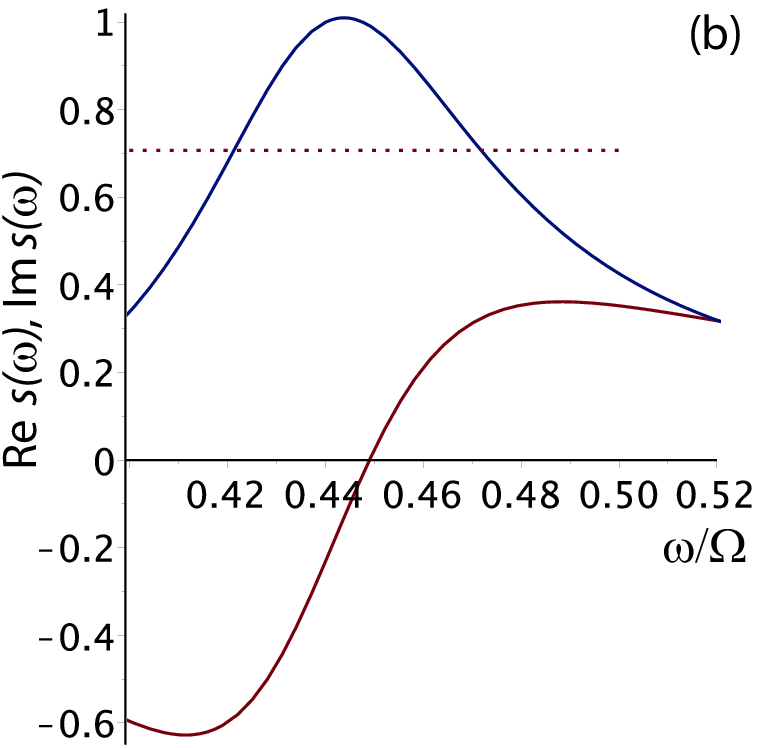}
\caption{(Color online) (a) Graphs of Re~$s$ (blue lines)
and Im~$s$ (red  lines) for $\eta = 2$, 5, and 10. (b) The same graphs for $\eta =5$. Dashed line shows the level of $1/\sqrt{2}$ for estimate of the shape of the curves. \label{tmp1}}
\end{figure}

Assuming the regime (b1) we can cast Eq.~(\ref{sigma1}) in the form
 $\sigma(\omega)\equiv \sigma_0  s(\omega/\Omega)$
where
\begin{equation}
\sigma_0 \equiv \frac{e^2n  \eta^2}{2m^*\omega_c}\,,  \quad
s(\tilde{\omega})=-2\frac{i \tilde{\omega}(1-i \tilde{\omega}^3)}
{\eta[(1-i \tilde{\omega}^3)^2- (\eta \tilde{\omega})^2]}\, .
\end{equation}
Graphs of $s(\omega/\Omega)$ for several $\eta$ are shown in
Fig.~\ref{tmp1}. This function is normalized in order to have its
maxima $\eta$-independent.
Unfortunately, the experimental data shown in Fig.~\ref{fig5}~(b) do
not provide an accurate structure of the maximum, and therefore do
not allow fitting the model with high accuracy. Assuming $\eta =5$
that gives approximately correct shape of the curves in
Fig.~\ref{fig5}~(b) and taking into account that the maximum of
$\sigma_1(\omega)$  occurring at $\omega_{\max}/2\pi \approx 85$ MHz
corresponds to $\omega /\Omega =0.44$ we conclude that $\Omega/2\pi
=\omega_{\max}/0.44\cdot 2\pi \approx 193$ MHz. The quantity
$\omega_p \equiv 0.44\Omega_{\max}=0.44\omega_{p0}^2/\omega_c$ plays
the role of pinning frequency in the magnetic
field~\cite{Fogler2000}.

The frequency $\omega_{p0}$ can then be determined as
$\omega_{p0}=\sqrt{\omega_c \Omega/\eta}$. Substituting $\eta=5$,
$\Omega/2\pi = 193$~MHz, $\omega_c/2\pi = 4\cdot 10^6$~MHz we obtain
$\omega_{p0}/2\pi =12.4\cdot 10^3$~MHz. Therefore, the regime (b) of
Eq.~(\ref{regimes}) is the case, as we expected. Estimating the WC
correlation length as $\xi =2 \pi c_t/\omega_{p0}$ where $c_t
\approx 4\cdot 10^6$~cm/s is the velocity of the WC transverse mode
for our electron density we obtain $\xi \approx 3 \cdot 10^{-4}$~cm
that is much larger both than the distance between the electrons and
the magnetic length $\ell_B =(c\hbar/eB)^{1/2}$. These inequalities
justify using the theory~\cite{Fogler2000} for estimates.

To summarize, from simultaneous measurements of attenuation and
velocity SAW propagating in proximity of 2DES in perpendicular
magnetic fields we have calculated both real and imaginary parts of
the complex AC conductance, $\sigma (\omega)$. Our analysis shows
that at low temperatures and at the filling factor of 0.20 the
electron system resides in the FQHE state. However, close to this
value, at $\nu=0.19$ and 0.21, the electron state can be interpreted
as a so-called weakly pinned Wigner crystal. When the temperature
(or the SAW intensity) increases the behavior of the complex
conductance can be understood as manifestation of WC melting. Our
conclusions support  the interpretation based on DC and microwave
measurements, as well as on previous acoustic measurements of only
real part of the AC conductance.

\section*{Acknowledgments}

I.L.D. is grateful for support from Russian Foundation for Basic
Research via grant 14-02-00232. The authors would like to thank E.
Palm, T. Murphy, J.-H. Park, and G. Jones for technical assistance.
NHMFL is supported by NSF Cooperative Agreement No. DMR-1157490 and
the State of Florida. The work at Princeton was partially funded by
the Gordon and Betty Moore Foundation through Grant GBMF2719, and by
the National Science Foundation MRSEC-DMR-0819860 at the Princeton
Center for Complex Materials.

\end{document}